\documentclass[pra,preprintnumbers,amsmath,amssymb,letterpaper,twocolumn,aps,superscriptaddress]{revtex4-1}

\usepackage{graphicx}
\usepackage{dcolumn}
\usepackage{bm}
\usepackage{epstopdf}
\usepackage{latexsym}
\usepackage{epsfig}
\usepackage{verbatim}
\usepackage{hyperref}
\usepackage{color}

\newcommand{\be}{\begin{equation}}
\newcommand{\ee}{\end{equation}}
\newcommand{\bea}{\begin{eqnarray}}
\newcommand{\eea}{\end{eqnarray}}

\newcommand{\req}[1]{Eq.~(\ref{#1})}
\newcommand{\reqs}[1]{Eqs.~(\ref{#1})}

\begin{document}

\title{Extreme High-Field Superconductivity in Thin Re Films} 

\author{F. N. Womack}
\affiliation{Department of Physics and Astronomy, Louisiana State University, Baton Rouge, Louisiana 70803, USA}
\author{D. P. Young}
\affiliation{Department of Physics and Astronomy, Louisiana State University, Baton Rouge, Louisiana 70803, USA}
\author{D. A. Browne}
\affiliation{Department of Physics and Astronomy, Louisiana State University, Baton Rouge, Louisiana 70803, USA}
\author{G. Catelani}
\affiliation{JARA Institute for Quantum Information (PGI-11), Forschungszentrum J\"ulich, 52425 J\"ulich, Germany}
\author{J. Jiang}
\affiliation{Department of Materials Science and Engineering, The University of Texas at Arlington, Arlington, Texas, USA}
\author{E. I. Meletis}
\affiliation{Department of Materials Science and Engineering, The University of Texas at Arlington, Arlington, Texas, USA}
\author{P. W. Adams}
\affiliation{Department of Physics and Astronomy, Louisiana State University, Baton Rouge, Louisiana 70803, USA}

\date{\today}

\begin{abstract}
We report the high-field superconducting properties of thin, disordered Re films via magneto-transport and tunneling density of states measurements.    Films with thicknesses in the range of 9 nm to 3 nm had normal state sheet resistances of $\sim0.2$ k$\Omega$ to $\sim1$ k$\Omega$ and corresponding transition temperatures in the range of 6 K to 3 K.  Tunneling spectra were consistent with those of a moderate coupling BCS superconductor.  Notwithstanding these unremarkable superconducting properties, the films exhibited an extraordinarily high upper critical field.  We estimate their zero-temperature $H_{c2}$ to be more than twice the Pauli limit.  Indeed, in 6~nm samples the estimated reduced critical field $H_{c2}/T_c\sim{5.6}\,$T/K is among the highest reported for any elemental superconductor.  Although the sheet resistances of the films were well below the quantum resistance $R_Q=h/4e^2$, their $H_{c2}$'s approached the theoretical upper limit of a strongly disordered superconductor for which $k_F\ell\sim1$.
\end{abstract}

\maketitle

\section{Introduction}

Over the last two decades superconductivity research has evolved along two separate but related paths.  The first is an extensive search and discovery effort aimed at identifying and characterizing superconducting phases that have novel order parameter symmetries and/or non-phonon coupling mechanisms.   Recent examples include systems with low to moderately high transition temperatures such as Sr$_2$RuO$_4$ \cite{SrRuO-a,SrRuO-b,SrRuO-c}, CeCoIn$_5$ \cite{CeCoIn,Yazdani}, Nb$_2$Pd$_{0.81}$S$_5$ \cite{NbPdS}, UTe$_2$ \cite{UTe2-a}, and the Fe-based arsenides \cite{FeAs-a,FeAs-b,FeAs-c}.  A second, although smaller, segment of the research effort is focused on novel quantum effects that can arise in BCS superconductors under certain conditions.  Examples include non-equilibrium dynamics in Zeeman-limited superconductors \cite{Adams1,Adams2}, disorder and correlation effects in thin film systems \cite{Valles1,Hebard,Klapwijk}, phase effects in systems having non-trivial multiply connected geometries\cite{Valles2,Shih,Adams3}, and, of course, quantum entanglement \cite{QC-a,QC-b}.   The results reported here fall into the latter category.  We present an experimental study of the extraordinary critical field behavior of thin, disordered Re films.  Notably, the films exhibit reduced critical fields as high as $H_{c2}/T_c\sim5.6\,$T/K, which is more than an order of magnitude greater than what is typical of elemental superconductors, and, in fact, one of the highest values reported for any superconductor.  We discuss the likely origins of the critical field enhancement and possible applications for which a large $H_{c2}$ would be advantageous.

\section{Experimental methods}

\begin{figure}[h]
\begin{center}
\includegraphics[width=.45\textwidth]{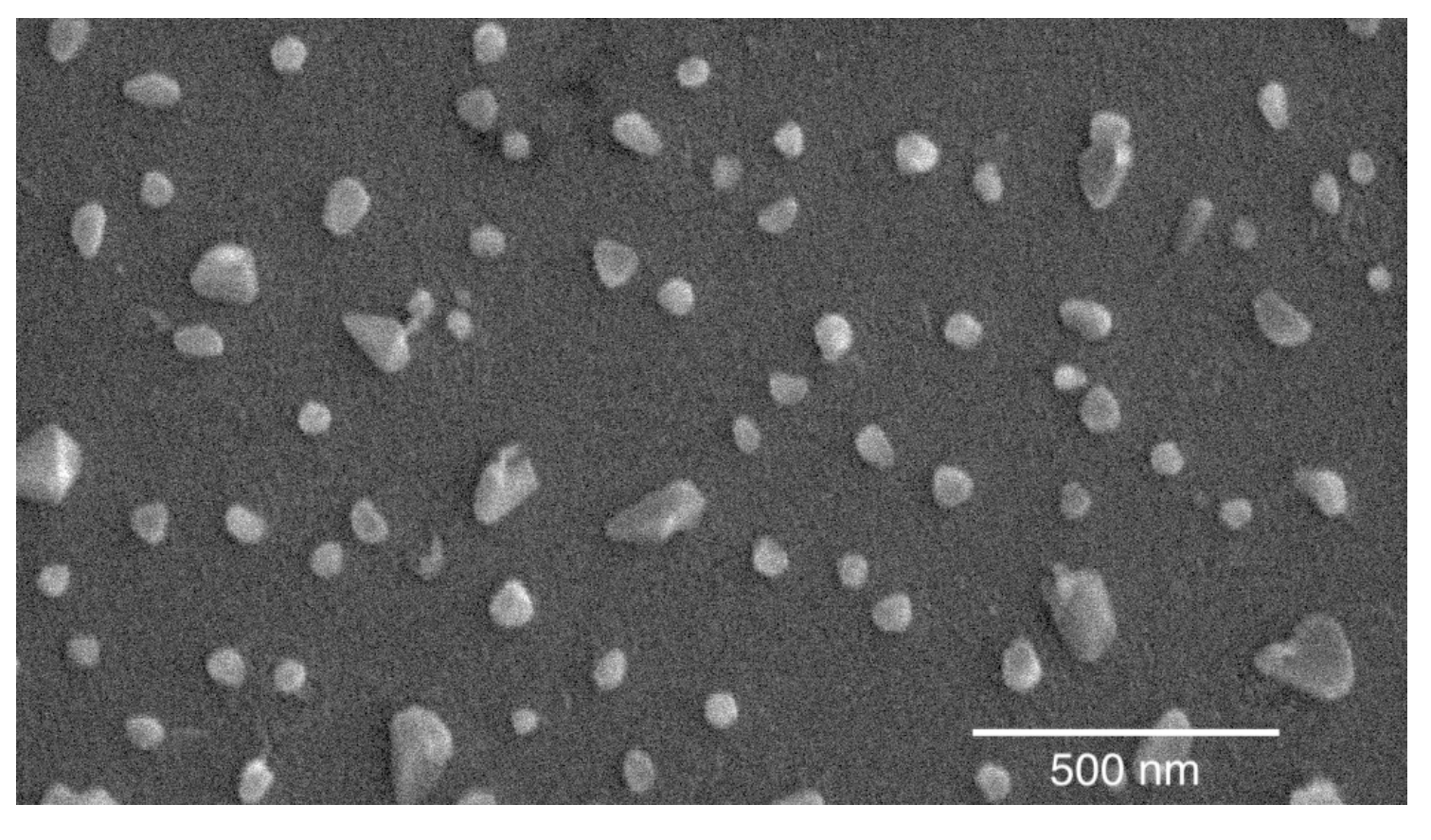}\end{center}
\caption{Scanning electron micrograph of a 10 nm-thick Re film deposited onto fire-polished glass at 84 K.}
\label{SEM}
\end{figure}

Rhenium films were formed by e-beam deposition from Re targets produced by arc-melting 99.9\% Re powder to form 2-3 mm diameter buttons.    The depositions were made onto fire-polished glass slides in a vacuum of $P<3\times10^{-7}$ Torr and a rate of $\sim0.5\,$\AA/s.  In order to minimize island formation in the films, the substrates were held at 84 K during the deposition, thus the films were effectively quench-condensed onto the cryogenic substrates.  Scanning electron micrographs (SEM) of a 10 nm-thick Re film revealed relatively large ($\sim$ 100 nm) Re particles scattered on what appeared to be a smooth, dense, amorphous Re base, see Fig.~\ref{SEM}. We believe these particles are the result of the ``spitting'' of Re droplets from the e-beam hearth.  Since their coverage is somewhat sparse we do not believe that they had a significant impact on the magneto-transport properties of the films. High resolution transmission electron microscopy (TEM) showed that the Re base is, in fact, granular on length scales of a few nm (see the next section). Planar tunnel junctions were formed by first depositing a counter-electrode composed of a non-superconducting Al alloy onto a glass substrate and then exposing the counter-electrode to atmosphere in order to produce a native oxide. A Re film was subsequently deposited so as to have partial overlap with the counter-electrode with the aluminum oxide serving as the tunnel barrier.  Magneto-transport measurements were made using a Quantum Design Physical Properties Measurement System having a maximum applied field of 9~T and a base temperature of 400~mK. The tunneling measurements were carried out using a standard 27~Hz 4-wire lock-in amplifier technique.

\section{Film Microstructure}

The microstructure of a 10 nm-thick Re film deposited onto fire-polished glass at 84 K was studied via scanning electron microscopy (SEM) and TEM. SEM analysis was performed on a 6 nm thick as-deposited Re film on fire-polished glass.  TEM analysis was performed on a 10 nm thick film that was mechanically transferred from fire-polished glass to a carbon coated TEM Cu grid. TEM images, selected-area electron diffraction (SAED) pattens, high-resolution (HR) TEM images and x-ray energy-dispersive spectroscopy (EDS) spectra were recorded in a Hitachi H-9500 electron microscope operating at 300 keV with a point resolution of 0.18 nm.

\begin{figure}[]
\includegraphics[width=.45\textwidth]{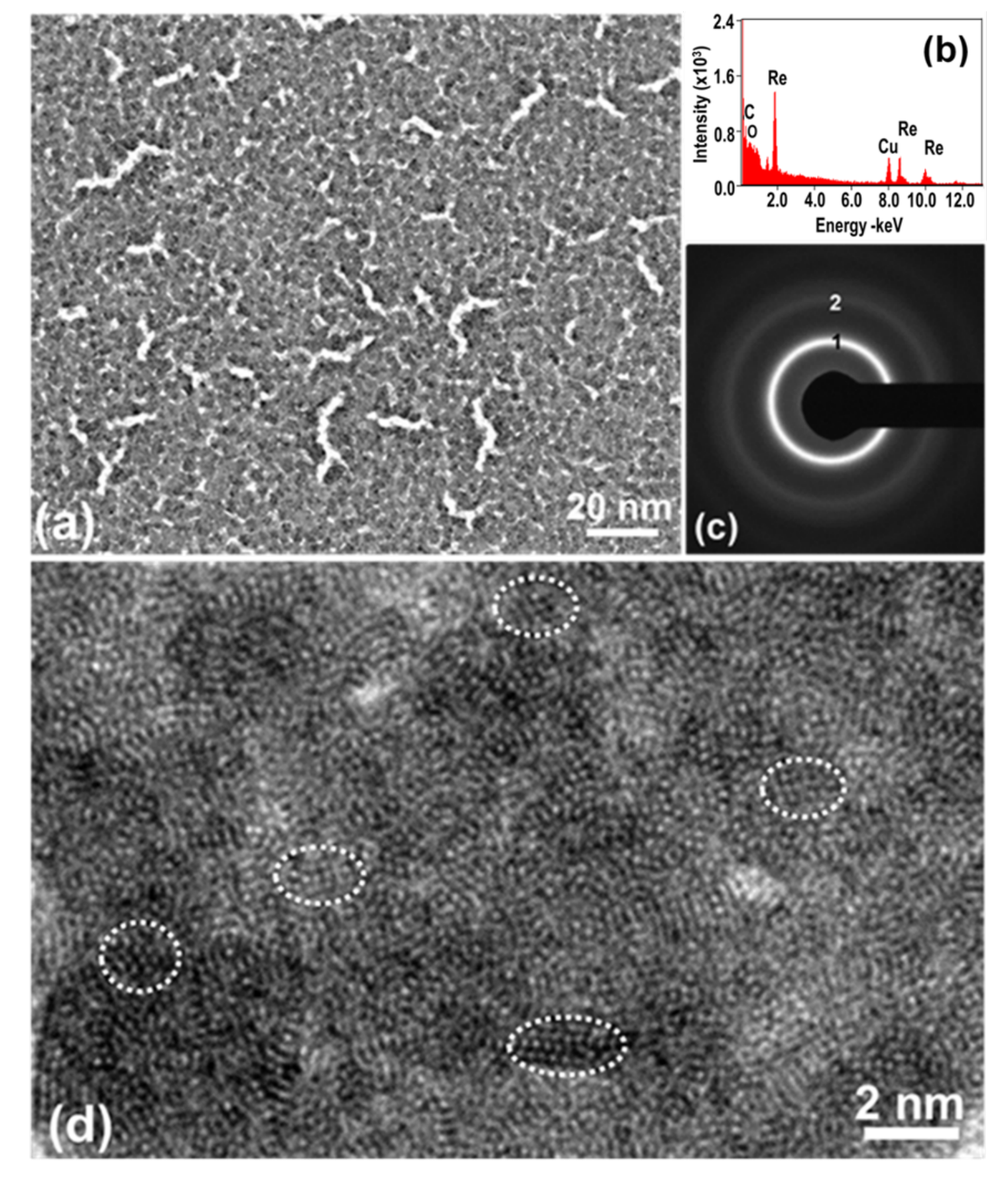}
\caption{(a) TEM image, (b) EDS spectrum, (c) SAED pattern and (d) HRTEM image of a 10 nm-thick Re film.}
\label{TEM}
\end{figure}

Figure~\ref{TEM}(a) presents a TEM image of a 10 nm film.  The film has a random granular structure with nano-grain sizes of few nm in diameter.  Figure~\ref{TEM}(b) presents an EDS spectrum of the film showing presence of Re peaks but no other elemental peaks except for that of the Cu in the TEM grid. The SAED pattern of the film exhibits diffused rings indicative of a short-range crystalline order within the nano-grains. Diffraction rings 1 and 2 in Fig.~\ref{TEM}(c) have an average d-spacing (measured from the middle point of the ring) of 2.17 \AA~and 1.28 \AA, respectively. We note that bulk Re has a hexagonal structure with a = 2.76 \AA~and c = 4.458 \AA~(P63/mmc). The (101)-Re plane has a d-spacing of 2.11 \AA, the (110)-Re plane has a spacing of 1.38 \AA, and the (103)-Re plane has a spacing of 1.26 \AA.  Therefore, diffraction ring 1 can be identified as the (101) plane of Re and the more diffuse ring 2 may correspond to an overlap of the (110) and (103) planes of Re. Figure~~\ref{TEM}(d) presents a HRTEM image of Re film in which the local arrangement Re atoms are revealed within an individual grain. A few characteristic crystallites having short range order are indicated by the marked regions.

\section{Background}

The quench-condensed Re films in this study are highly disordered as evidenced by their sheet resistances and the transmission electron micrographs.  The effects of disorder on two-dimensional (2D) BCS superconductivity has been the subject of intense investigation for more than 30 years now \cite{Valles1,Valles2,Hebard,Klapwijk}.  Early studies suggested that in homogeneously disordered 2D systems the underlying repulsive Coulomb correlations are enhanced by disorder \cite{Butko}.  This serves to suppress the transition temperature, and at a relatively well-defined critical level of disorder the superconducting phase is lost.  This disorder threshold is quantified by the quantum resistance $R_Q=h/4e^2=6.4$ k$\Omega$/sq, where $h$ is Planck's constant and $e$ is the electron charge.   Relatively recent studies have revealed a more complex picture in which disorder produces effects that arise from both single-electron and many-body quantum processes.  In terms of the complex superconducting order parameter $\Psi=\Delta_0e^{i\phi}$, the two limiting pathways to the complete destruction of the superconducting phase are: the suppression of the amplitude $\Delta_0$ or suppression of the phase stiffness $\phi$~\cite{Klapwijk}.  In the present study the sheet resistance of the samples are well below $R_Q$, so the superconducting phase remains relatively robust although it is clear from Fig.~\ref{R-T} that $T_c$ is somewhat suppressed in the thinner films.  Likely, samples with thicknesses substantially below 3 nm will have sheet resistances near or above $R_Q$ and these will presumably be at the threshold of a superconductor-insulator transition, but these considerations are beyond the scope of the present study.

In general, the critical field of a thin film superconductor has both an orbital and a Zeeman component.  The latter originates from the Zeeman splitting of the conduction electrons \cite{Fulde}.   In most circumstances, however, the orbital response of the superconductor dominates its critical field behavior in the sense that the Zeeman critical field can be an order of magnitude larger that its orbital counterpart.  This is particularly true in high spin-orbit (SO) scattering superconductors such as Nb and Pb due to the fact that even relatively modest SO scattering rates can dramatically quench the Zeeman response~\cite{SO}.  In the absence of SO scattering one can realize a purely Zeeman-mediated, first-order critical field transition~\cite{Adams1} by applying a magnetic field parallel to the surface of a thin film superconductor. The corresponding parallel critical field is given by the Clogston-Chandrasekhar equation~\cite{CC},
\be
 H_{cp}(0)=\frac{\sqrt2\Delta_0}{g\mu_{\rm B}},
 \label{Hcp}
 \ee
where $\Delta_0$ is the zero temperature - zero field gap energy, $\mu_{\rm B}$ is the Bohr magneton, and $g\sim2$ is the Land\'{e} g-factor.

The critical field of \req{Hcp} is known as the Pauli limit and represents the upper limiting critical field of a low SO superconductor.  In practice, the film thickness $d$ should be smaller than the penetration depth, $d \lesssim \lambda$,
in which case the magnetic field fully penetrates the sample, and also smaller than the coherence length, $d \lesssim \xi$.  Under these conditions the Zeeman-limited critical field transition occurs near the Pauli limit given by \req{Hcp}.  In contrast, if the field is applied perpendicularly to a thin superconducting film, an array of quantized vortices is induced whose area density is roughly proportional to the field strength.   The perpendicular critical field is reached when the vortex density approaches the point where the vortex cores begin to overlap,
\be
H_{c2}(0)=\frac{\Phi_0}{2\pi\xi^2},
\label{Hc2}
\ee
where $\Phi_0$ is the flux quantum and $\xi$ is the Pippard coherence length~\cite{Tinkham}.  The films in this study where quench-condensed onto cold substrates and thus were highly disordered.   Consequently their coherence length was a function of the mean free path $\ell$, $\xi=\sqrt{\xi_0\ell}$, where $\xi_0$ is the BCS coherence length.  The latter can be expressed in terms of the Fermi wave vector $k_F$ \cite{Tinkham},
\be
\xi_0=\frac{\hbar^2k_F}{\pi m\Delta_0},
\label{xi0}
\ee
where $m\simeq 2.1m_e$ is the effective mass estimated from band-structure calculations (see \cite{Re-fermi} and Appendix~\ref{app:rebs}), and $m_e$ the bare electron mass.   Combining \reqs{Hc2} and (\ref{xi0}) we obtain an expression for the perpendicular critical field in the strong-disorder limit $k_F\ell\rightarrow1$,
\be
H^{\rm max}_{c2}(0)=\frac{\Phi_0m\Delta_0}{2\hbar^2}.
\label{Hc2max}
\ee
Equation (\ref{Hc2max}) gives a rough estimate of the disordered-enhanced, perpendicular critical field of a dirty superconductor.  In practice this upper limit in Re is about 28~T per meV of gap.  We note that this is somewhat larger than the Pauli critical field, which is about 12~T per meV of gap; in fact, comparing Eq.~(\ref{Hcp}) with $g=2$ to Eq.~(\ref{Hc2max}), we find $H^{\rm max}_{c2}= H_{cp} \pi m/2\sqrt{2}m_e$.

Equations (\ref{Hcp}), (\ref{Hc2}), and (\ref{Hc2max}) represent low temperature critical fields $T\ll T_c$.  Unfortunately, the critical fields of the Re films in this study were well beyond the 9 T limit of our measuring system.  Nevertheless, reasonably accurate estimates of the both $H_{c\parallel}(0)$ and $H_{c2}(0)$ can be extracted from the temperature dependence of the respective critical fields near $T_c$.  We use the Werthamer-Helfand-Hohenberg formula \cite{WHH} to obtain the $T=0$ orbital critical field,
\be
H_{c2}(0)=-0.693\left(\frac{dH_{c2}}{dT}\right)_{T_c}\!\times T_c.
\label{WHH}
\ee
Similarly, the parallel critical field can be estimated via the following (see Appendix~\ref{app:theo}),
\be
H^2_{c\parallel}(0)=-0.693\left(\frac{dH^2_{c\parallel}}{dT}\right)_{T_c}\!\times T_c.
\label{Hc||}
\ee
Equation (\ref{Hc||}) accounts for both the Zeeman and orbital responses of the superconductor.  Of course, the orbital response of a thin film to parallel field is greatly suppressed if $d\ll\xi$, but it is not zero.   The Zeeman response is independent of geometry but is inhibited by SO scattering.  As a consequence of these effects, the measured $H_{c\parallel}$ can be more than an order of magnitude greater than $H_{cp}$.  Since the intrinsic SO scattering rate is proportional to $Z^4$ \cite{SO}, where $Z$ is the atomic number of the element, all but the lightest elements (such as Al and Be) have significant SO enhancements of $H_{c\parallel}$.  As we will show below, Re films in this study also exhibited parallel critical fields well above $H_{cp}$ due to their high intrinsic SO scattering rate. An alternative explanation in terms of the Rashba spin-orbit effect is also possible, as discussed in Appendix~\ref{app:theo}, but it does not modify our conclusions.

\section{Results and discussion}
\label{sec:results}

\begin{figure}
\begin{flushleft}
\includegraphics[width=.44\textwidth]{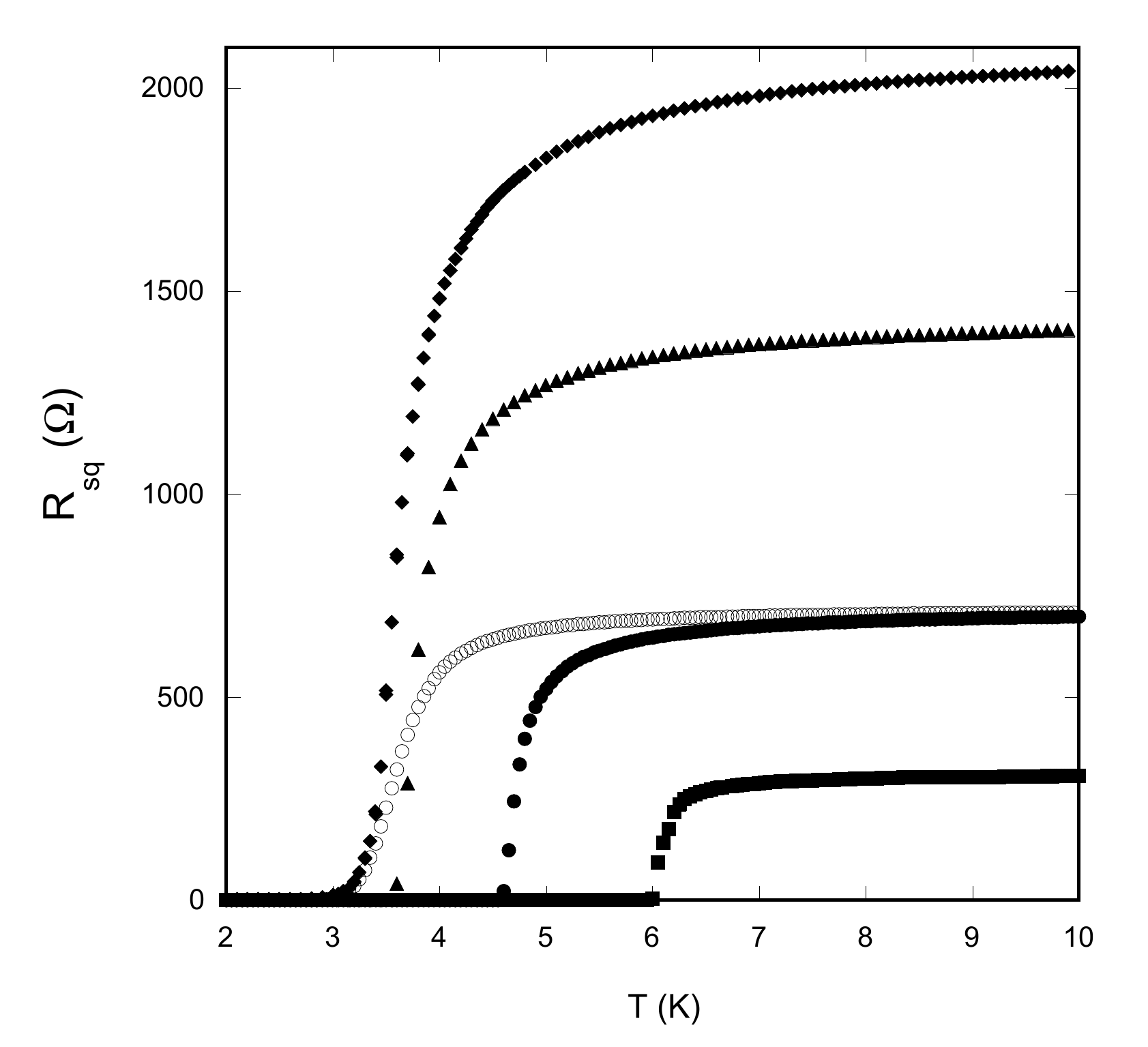}\end{flushleft}
\caption{Sheet resistance as a function of temperature showing superconducting transitions in a 2.5 nm (diamonds), 3 nm (triangles), 6 nm (filled circles), and a 9 nm thick (squares) Re film.  The open circles represent the superconducting transition of the 6 nm film in the presence of a 9 T perpendicular magnetic field.}
\label{R-T}
\end{figure}

Shown in Fig.~\ref{R-T} are the zero-field superconducting transitions of a 2.5~nm, 3~nm, 6~nm and 9~nm Re film.  Note that the sheet resistances of all films were well below the quantum resistance $R_Q$, indicating that, although the films were significantly disordered, they were not near the superconductor-insulator transition \cite{Klapwijk}.  The transition temperatures in Fig.~\ref{R-T} are considerably higher than that of bulk Re ($T_c=1.7\,$K), but it has been known since the mid 1950's that Re has a compliant $T_c$ which can be non-perturbatively enhanced over its bulk value by pressure, strain, and/or milling~\cite{Re-a,Re-b,Re-c,Re-d}.  In this respect it is not surprising that our films, which presumably have large lateral strains by virtue of the deposition technique, also exhibit enhanced $T_c$'s over the bulk value.  However, the films are also highly disordered, which presumably negatively impacts their $T_c$'s.  In the thickness range of this study $d=2\rightarrow9\,$nm, the transition temperature increased approximately linearly with $d$ from $T_c\approx3\rightarrow6\,$K.

\begin{figure}
\begin{flushleft}
\includegraphics[width=.44\textwidth]{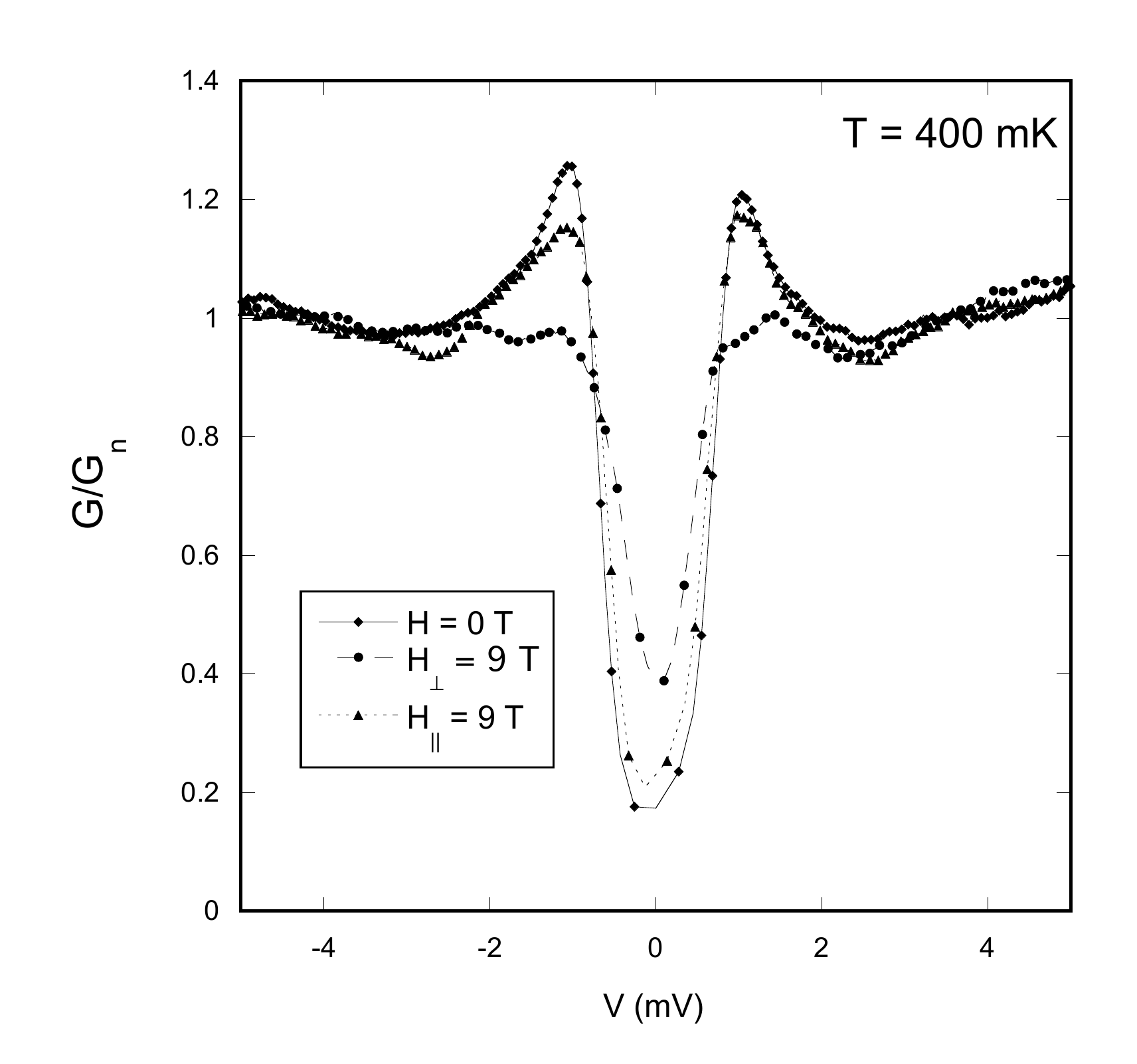}\end{flushleft}
\caption{Normalized tunneling conductance as a function of bias voltage in the superconducting phase of a $d=6\,$nm Re film.  The coherence peaks are associated with a gap energy $\Delta_0\sim1\,$meV.}
\label{G-V}
\end{figure}

Shown in Fig.~\ref{G-V} are tunneling conductance spectra taken on a $d=6\,$nm Re film.  At low temperatures the tunneling conductance $G$ is proportional to the single-particle density of states (DoS)~\cite{Tinkham}.  The bias voltage is relative to the Fermi energy, and the conductances have been normalized by the conductance at 4~mV.  The zero-field spectrum shows well-defined coherence peaks associated with a gap energy $\Delta_0=1.0\,$meV. The ratio of the gap energy to transition temperature $\Delta_0/k_{\rm B}T_c\sim2.4$ is moderately larger than the expected BCS value of 1.76.  A similar disorder-induced enhancement of the coupling strength is also observed in disordered Al films \cite{Al,Scheffler}.  The non-zero conductance near $V=0$ is partially an artifact of the finite input impedance of the lock-in amplifier, but overall, the spectrum is consistent with that of a disordered BCS superconductor.  Note that the application of a 9~T perpendicular field suppresses the coherence peaks but does not completely extinguish the gap.  Indeed, the excess zero bias conductance of the perpendicular field trace is associated with the cores of the induced vortices.  The parallel field spectra is little changed from its zero-field counterpart, indicating that the film is too thin to accommodate vortices.

\begin{figure}
\begin{flushleft}
\includegraphics[width=.44\textwidth]{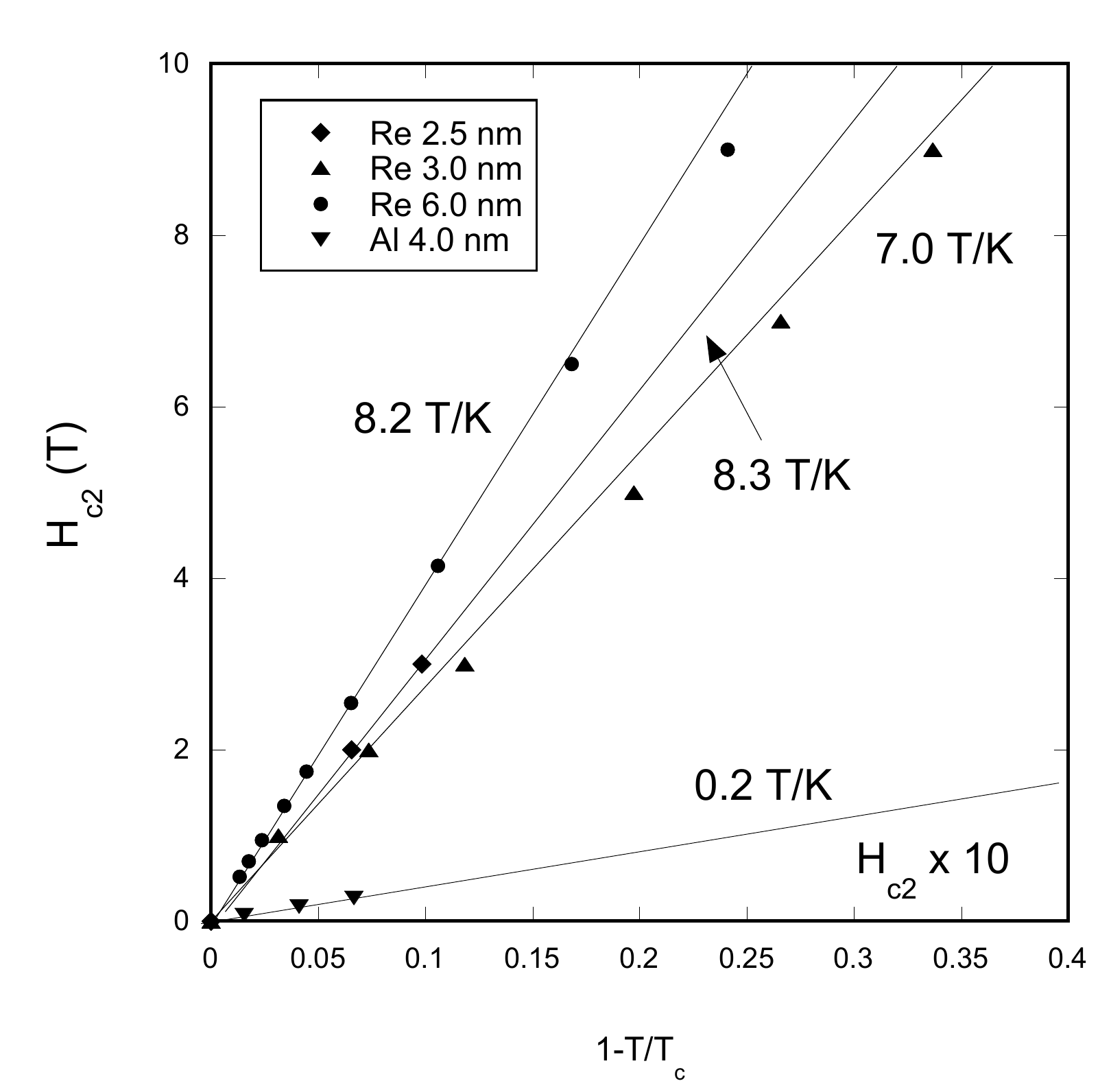}\end{flushleft}
\caption{Perpendicular critical field as a function of reduced temperature near $T_c$ for several Re films of varying thickness.   For comparison, the inverted triangles represent the critical field $(\times10)$ of a $d=4\,$nm Al film that was capped with 0.3~nm Au to induce spin-orbit (SO) scattering. The solid lines are linear least-squares fits to the data below 3 T.  The values of $(dH_{c2}/dT)_{T_c}$ are indicated in the panel.}
\label{OCF}
\end{figure}

As is evident in the 6~nm data shown in Fig.~\ref{R-T}, the disordered Re films have extraordinarily high critical fields, although their zero-field superconducting characteristics seem quite conventional.  Note that a 9~T perpendicular field, which is the upper limit of our system, shifted the transition temperature of the 6~nm film by only 20\%!  In order to estimate the $T=0$ critical fields, we have employed \reqs{WHH} and (\ref{Hc||}).   This was done by either measuring the critical field at temperatures near $T_c$ or by measuring the transition temperature in the presence of a static applied field, see Appendix~\ref{app:cft}.  In either case the transition was define by the temperature/field at which the resistance reached $1/2$ of its normal state value. These two methods gave equivalent results but sweeping temperature in a constant magnetic field proved to be more expedient.

In Fig.\ \ref{OCF} we plot $H_{c2}$ as a function of reduced temperature near $T_c$. Note that the perpendicular critical field increases linearly with decreasing temperature suggesting that it is dominated by the orbital contribution.  The solid lines are a linear least-squares fit to the data below 3 T, and their slopes provide an indirect measure of $H_{c2}(0)$ via \req{WHH}.  For the 6~nm Re film in Fig.~\ref{R-T} we get $(dH_{c2}/dT)_{T_c}=8.2\,$T/K.  Using \req{WHH} and neglecting the Zeeman response, we estimate that $H_{c2}(0)\approx27$ T for this sample.  This is quite close to $H^{\rm max}_{c2}\approx28\,$T from \req{Hc2max}, which implies that the films are in the strong disorder limit of $k_F\ell\sim1$.

The corresponding reduced critical field of the 6~nm film is also very large $h\equiv H_{c2}(0)/T_c\sim5.6\,$T/K. In fact, this is among the highest reduced fields reported in the literature.  Typically, $h<1$ in elemental films, for example: $h\sim1$ T/K in highly disordered granular Pb films \cite{Valles3}, $h\sim0.15$ in thin amorphous Be films \cite{Be-h}, and $h\sim.3$ in ultra-thin crystalline Pb films \cite{PbX}.  For bulk systems, Chevrel-phase PbMo$_6$S$_8$ \cite{PbMoS} has $T_c=13.3\,$K and $H_{c2}(0)\sim60\,$T, giving $h=4.5\,$T/K.  We note that the critical field slope near $T_c$ of PbMo$_6$S$_8$ is 6.4~T/K from which \req{WHH} predicts an extrapolated critical field of 59 T, in good agreement with the measured value. The Re film reduced critical field is also comparable to the $b$-axis critical field of the highly anisotropic chalcogenide Nb$_2$Pd$_0.81$S$_5$ \cite{NbPdS}, $h=5.6\,$T/K but is not as large as that of the spin-triplet superconductor UTe$_2$ \cite{UTe2-a,UTe2-b}, $h\sim20\,$T/K.  All of these reduced critical fields are, in fact, substantially larger than that of one of the most important high field superconductors Nb$_3$Sn \cite{NbSn}, $h=1.7\,$T/K.

\begin{figure}
\begin{flushleft}
\includegraphics[width=.44\textwidth]{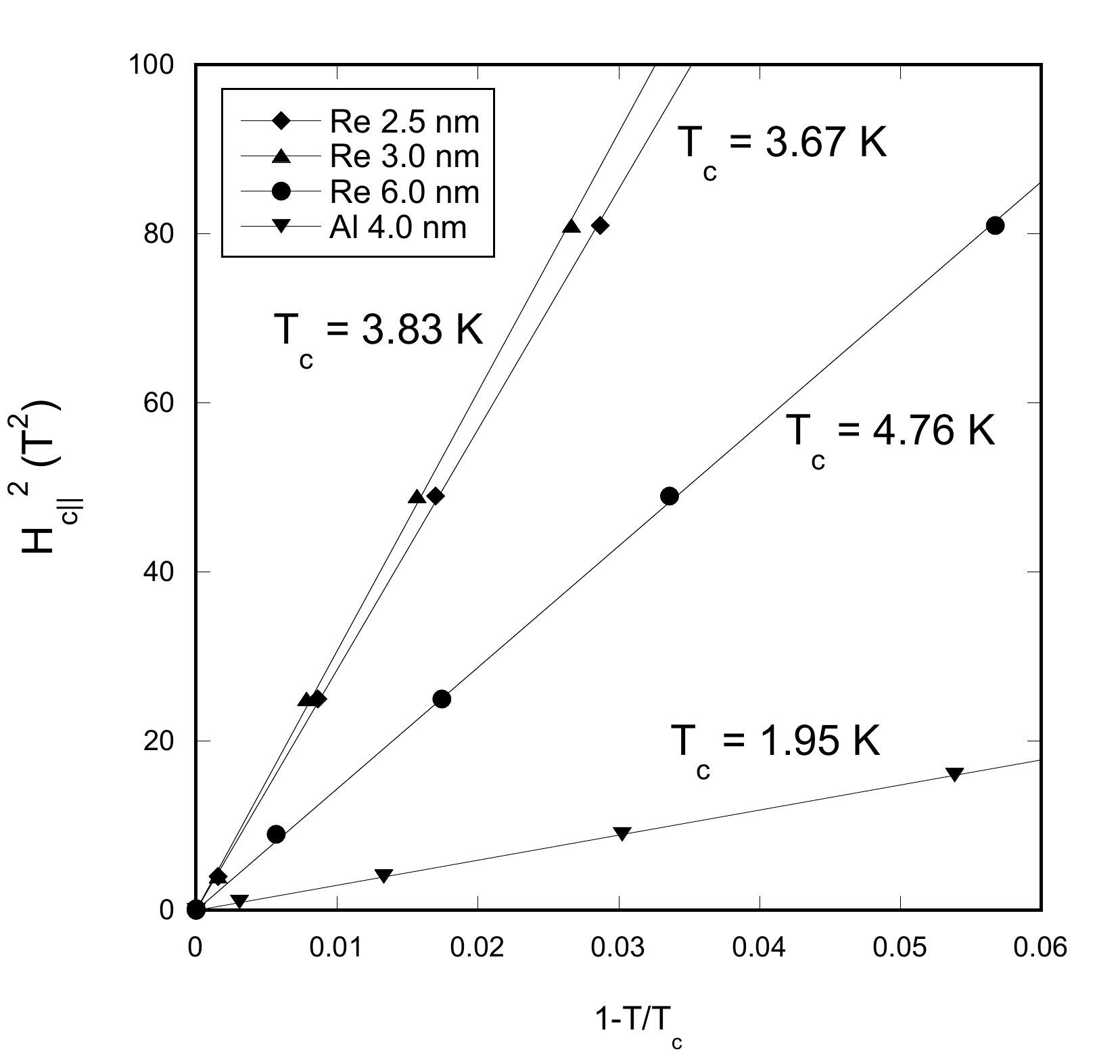}\end{flushleft}
\caption{Parallel critical fields of the samples from Fig.~\ref{OCF} as a function of reduced temperature.  The solid lines are linear least-squares fits to the data.}
\label{PCF}
\end{figure}

For comparison we have included critical field measurements in Fig.~\ref{OCF} of a 4~nm thick Al film that was capped with 0.3~nm of Au to induce SO scattering~\cite{TM}.  The transition temperature of the Al film was roughly half that of the 6~nm Re film.  The magnitude of the Al critical field $H_{c2}(0)=0.27\,$T is typical of what is seen in thin elemental films.  It is striking that $dH_{c2}/dT$ of the Re film is 40 times larger than that of the Al film in Fig.~\ref{OCF}, but their respective sheet resistances only differ by a factor 4.  Since $H_{c2}$ scales as $\Delta_0/\ell\sim T_c/\ell$, one would expect the Al film critical field to be about 1/8 that of the Re film when, in fact, they differ by a factor of 100.  The corresponding coherence lengths can be calculated from \req{Hc2}, and they differ by an order of magnitude: $\xi_{\rm Re}=3.5\,$nm and  $\xi_{\rm Al}=35\,$nm.  Interestingly, the Re coherence length is comparable to the scale of the structural granularity of the films as revealed by TEM analysis.

We have also measured the parallel field response of the films.  In Fig.~\ref{PCF} we plot the square of the parallel critical field as a function of reduced temperature near $T_c$.  As suggested by \req{Hc||}, $H_{c\parallel}^2$ should be linear in temperature.  The solid lines in Fig.~\ref{PCF} represent least-squares fits to the data from which we can obtain an estimate of $H_{c\parallel}(0)$ as per \req{Hc||}.  The corresponding low temperature parallel critical fields for the Re 2.5~nm, Re 3~nm, Re 6~nm, and Al 4~nm films are estimated to be 46.1~T, 31.5~T, and 14.4~T, respectively.  Since the parallel critical field is predominantly mediated by the spin response of the system, $\ell$ does not play as large a role as it does in perpendicular field, and  $H_{c\parallel}$ is primarily determined by $\Delta_0$ and the SO scattering rate. Interestingly, the three films have reduced critical fields $h_{c\parallel}=H_{c\parallel}/T_c\sim 7$-$12\,$~T/K, which are significantly larger than the reduced field in the absence of SO scattering, $h_{cp}\sim1.8\,$T/K. The parallel field behavior of the Re films in Fig.~\ref{PCF} is very similar to that of the Al both in temperature dependence and magnitude.  But there is a profound difference in the ratio of the parallel critical field to the perpendicular critical field of the two systems.  For the Re films $H_{c\parallel}/H_{c2}\sim1.2$ to 2.7 but, in contrast, $H_{c\parallel}/H_{c2}\sim50$ for the Al film.

The $T_c$ enhancement in our quenched-condensed Re films is similar in magnitude to what is seen in much thicker electroplated Re films ($\sim30\,$nm)~\cite{Re-c}.  This suggests that the nano-morphology or perhaps the strain-induced changes in the bulk lattice constant play an important role in determining $T_c$ \cite{Re-d}.  Of course, the high perpendicular critical field of the films may, in part, be explained by their disorder. The band structure of bulk Re gives $v_F\sim8\times10^5\,$m/s, see Appendix~\ref{app:rebs}. Combining $v_F$ with the measured gap $\Delta_0=1\,$meV points to a BCS coherence length  $\xi_0\sim170\,$nm.  This, in turn, implies that the mean-free-path of the film would need to be $\ell=\xi^2/\xi_0\sim1\,$\AA \, which is of the order of the inter-atomic spacing.  This seems to be an unreasonably short $\ell$, since it suggests that the films are deep in the Anderson localization limit. However, they do not exhibit correlated insulator behavior~\cite{QM} that is often associated with strong disorder.  This is perhaps due to the fact that the density of states of Re is extraordinarily high, about twice that of Cu.

\section{Summary}

In summary, we report very large perpendicular critical fields in Re films quenched condensed onto liquid nitrogen-cooled substrates.  There continues to be a substantial interest in liquid-helium temperature superconductors that can be used in quantum information/computation technologies~\cite{houck,pop1,bylander}.  This is particularly true of thin-film superconductors that are: easily deposited, resistant to oxidation, have low resistivity, and/or are compatible with high magnetic fields~\cite{leodc,pop2}.  Rhenium offers a compelling alternative to Al in devices such as superconducting resonators and microwave circuits~\cite{Re-d,Gilles}.  Extending the present studies to direct ultra-high field measurements of the low-temperature magnetotransport and tunneling density of states properties of the films should prove enlightening.

\acknowledgments

The magneto-transport measurements were performed by P.W.A.\ and F.N.W with the support of the U.S. Department of Energy, Office of Science, Basic Energy Sciences, under Award No.\ DE-FG02-07ER46420.    Film synthesis was carried out by  D.P.Y.\ and F.N.W.  D.P.Y. acknowledges support from the NSF under Award No.\ DMR-1904636. The TEM analysis was performed by J.J. and E.I.M. The theoretical critical field analysis was carried out by G.C. Band structure calculations were performed by D.A.B.

\appendix

\section{Rhenium band structure}
\label{app:rebs}

The band structure of Re was calculated using the ab-initio LAPW method
implemented in the WIEN2k software~\cite{WIEN2k} using a PBE functional~\cite{PBE}
and included the spin-orbit interaction.  Lattice constants were chosen
as 5.211 and 8.404 Bohr (2.758 and 4.447~\AA).  The Re muffin tin
radius was chosen as 2.5 Bohr.  In addition to the LAPW orbitals, an
additional relativistic $p_{3/2}$ orbital \cite{RO} at -3.08 Ryd below $E_F$ was
added to improve the basis set.  The plane-wave cutoff was varied from
R*G$_{\rm max}$ = 6 to 8 with no significant change observed in the band
structure.  For integration over the Brillouin zone, grids of 10,000 to
30,000 points were used.  To calculate the Fermi surfaces a 100$^3$ grid
was used.

\begin{figure}[!tb]
\includegraphics[width=.4\textwidth]{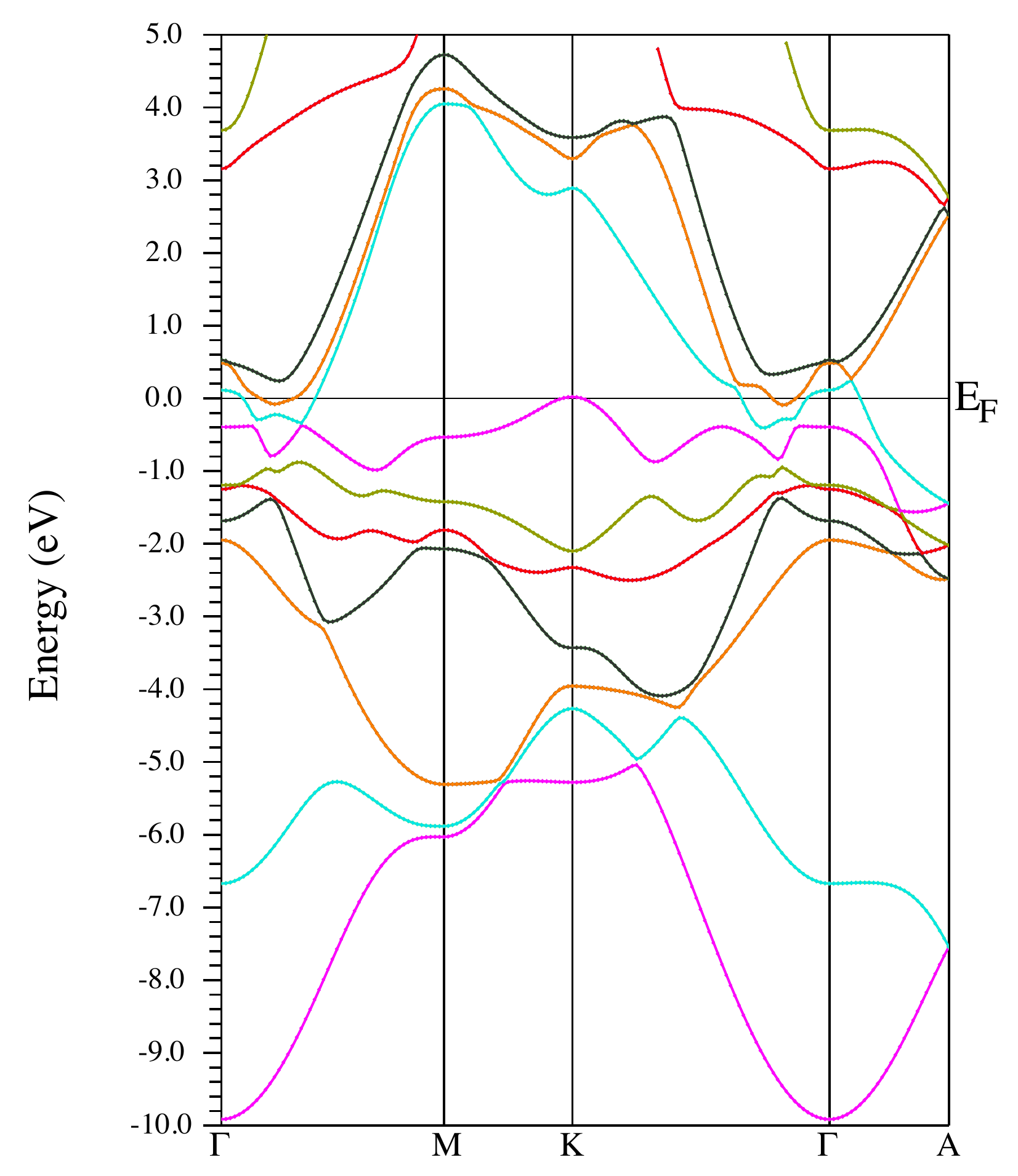}
\caption{(Color online) Relativistic energy bands of rhenium, including the effects of spin-orbit coupling.}
\label{fig:BS}
\end{figure}

The Fermi surfaces were very similar to those seen in Mattheiss'
work~\cite{Re-fermi}.  A set of 5 sheets cross the Fermi level.  The density of
states was found to be 0.66 states/eV/atom, similar to what Matthiess'
found.   Assuming 2 valence electrons per Re, the band effective mass
is $2.1m_e$.  The mean Fermi velocity was found on each sheet, and it
varied from 2.7$\times10^5$ to 8.6$\times10^5\,$m/s among the sheets.  Using an expected gap
$\Delta_0$ of 1.05~mV, that gives a typical coherence length
$\xi_0=\hbar v_F/\pi \Delta_0$ of 55 to 170 nm, depending on the
band.

\section{Estimating the zero-temperature critical field}
\label{app:theo}

\subsection{Theoretical background}
\label{sec:theo}

Generically, a second-order phase transition between superconducting and normal states at temperature $T<T_c$ takes place when the strength of a pair-breaking mechanism reaches a critical value. If the energy scale associated with such a mechanism is denoted by $\alpha$ (which is proportional to the pair-breaking rate), then the critical value is found in many cases, but not all, by solving for $\alpha$ as a function of $T$ using the following equation (cf. Sec.~10.2 of Ref.~\cite{Tinkham})
\begin{equation}\label{eq:pbe}
    \ln\frac{T}{T_c} = \Psi\left(\frac12\right) - \Psi\left(\frac12+\frac{\alpha}{2\pi k_B T}\right)
\end{equation}
where $\Psi$ is the digamma function. As $T \to T_c$, we expect $\alpha\to0$, since at $T=T_c$ the transition takes place in the absence of the additional pair-breaking mechanism, and by a Taylor expansion of the right hand side of \req{eq:pbe} we find
\begin{equation}\label{eq:atc}
    t = \Psi'\left(\frac12\right)\frac{\alpha(T)}{2\pi k_B T_c},
\end{equation}
where  $\Psi'(1/2) = \pi^2/2$. Conversely, as $T\to0$ we can use the identity $\Psi(1/2) = -2\ln2 - \gamma_E$, where $\gamma_E\simeq0.577$ is Euler's constant, and the asymptotic formula
\begin{equation}
    \Psi\left(\frac12+x\right) \simeq \ln x + \mathcal{O}\left(x^{-2}\right)
\end{equation}
to find
\begin{equation}\label{eq:a0}
    \ln \frac{\alpha(0)}{2\pi k_B T_c} = - 2\ln 2-\gamma_E.
\end{equation}
Using the fact that the zero temperature gap and the critical temperature are related as $\Delta_0 = \pi e^{-\gamma_E} k_B T_c \approx 1.76 k_B T_c$, we can rewrite this equation as
\begin{equation}\label{eq:a0D0}
    \alpha(0) = \frac{\Delta_0}{2}.
\end{equation}

Differentiating \req{eq:atc} with respect to temperature and comparing the result to \req{eq:a0}, we can relate the temperature derivative of $\alpha$ near $T_c$ to the value of $\alpha$ at zero temperature:
\begin{equation}\label{eq:a0ap}
    \alpha(0) = -\frac{\pi^2}{8}e^{-\gamma_E} \frac{d\alpha}{dT}\bigg|_{T_c} T_c
\end{equation}
where for the numerical coefficient we have $\pi^2e^{-\gamma_E}/8 \approx0.693$.

The above approach is applicable to various pair-breaking mechanisms. For the orbital effects of perpendicular and parallel magnetic fields of magnitude $H$ applied to a disordered film we have, respectively~\cite{Tinkham},
\begin{equation}\label{eq:aTink}
    \alpha_\perp = D e H \, , \qquad \alpha_\parallel = \frac{D}{6\hbar} \left(eHd\right)^2
\end{equation}
where $D$ is the diffusion constant and $d$ the film thickness. Note that using the expression for $\alpha_\perp$ in Eq.~(\ref{eq:a0D0}) we obtain the familiar result
\begin{equation}\label{eq:hc20}
    H_{c2} (0) = \frac{\Phi_0}{2\pi \xi^2}
\end{equation}
with $\Phi_0 = \pi\hbar/e$ the magnetic flux quantum and $\xi =\sqrt{\hbar D/\Delta_0}$ the zero-temperature coherence length.

Generically, independent pair-breaking mechanisms can be added together, similar to ``Matthiessen's rule'' of adding together scattering rates. In fact, one can in this way derive Tinkham's formula~\cite{Tinkham}
\begin{equation}
\frac{H_c(\theta)\cos\theta}{H_{c\perp}(T)} + \left[\frac{H_c(\theta)\sin\theta}{H_{c\parallel}(T)}\right]^2 = 1,
\end{equation}
for the angular dependence of the critical field of a thin film, where $\theta$ is the angle between field and normal to the film: according to Eq.~(\ref{eq:a0}), at any given temperature we have that $\alpha_\perp+\alpha_\parallel = DeH_c(\theta)\cos\theta + D(ed)^2[H_c(\theta)\sin\theta]^2/6\hbar$ is constant, and that constant can be written as both $DeH_{c\perp}(T)$ and $D(ed)^2[H_{c\parallel}(T)]^2/6\hbar$.

An important case in which the simple summation of pair-breaking strengths is not valid arises when considering both orbital effects and Zeeman splitting due to a magnetic field; a full treatment for the case of parallel field including orbital effect, Zeeman splitting, Fermi-liquid renormalization of the spin susceptibility, and spin-orbit scattering can be found in Ref.~\cite{Alexander}, while we can employ the results of Ref.~\cite{Maki} for the perpendicular field case (see also Ref.~\cite{WHH}).
Fortunately, if spin-orbit scattering is sufficiently strong (in a way to be specified below), then one can still use the ``Matthiessen's rule'' approach by modifying the pair-breaking energy $\alpha$. For parallel field, we have
\begin{equation}\label{eq:ap_so}
    \alpha_\parallel = \Delta_0 \left(c+\frac{1}{2b}\right) \tilde{h}^2
\end{equation}
where $\tilde{h}= \mu_B H/\Delta_0$ and parameters $c$ and $b$ account for orbital effect and spin-orbit scattering
\begin{equation}\label{eq:parpar}
    c= \frac{D \Delta_0}{6\hbar} \left(\frac{ed}{\mu_B}\right)^2 \, \qquad b = \frac{\hbar}{3\Delta_0 \tau_{so}},
\end{equation}
where $\tau_{so}$ is the spin-orbit scattering time.

For Eq.~(\ref{eq:ap_so}) to be valid, we need $b\gg \tilde{h}$; in fact, Eq.~(\ref{eq:ap_so}) can be obtained by taking the large $b$ limit in the more general formulas (valid for any values of $c$, $b$, and $\tilde{h}$) given in Ref.~\cite{Alexander} (we neglect the possible Fermi-liquid renormalization of the spin susceptibility, since in the regime studied here it can be absorbed in a redefinition of $b$). Using Eq.~(\ref{eq:ap_so}) in Eq.~(\ref{eq:atc}) we find that near the critical temperature $H_{c\parallel}^2 \propto t$. If $c < 1/2b$ and the film is very thin, then the parallel critical field is limited by the Zeeman effect, but at a field well above the Pauli limit.  Interestingly, if a film is only moderately thin ($c > 1/2b$) then the parallel critical field can be limited by the orbital effect, but again at a field well above the Pauli limit. Substituting Eq.~(\ref{eq:ap_so}) into Eq.~(\ref{eq:a0ap}), we obtain Eq.~(\ref{Hc||}), which is valid whether the parallel critical field is Zeeman-limited or orbital-limited, so long as $b\gg \tilde{h}$. As an aside, we remind that in the opposite case of negligible spin-orbit scattering, the Zeeman splitting dominates over the orbital effect of the field if $c\ll 1$, a condition that can be written also as $d\ll \xi/k_F \ell$; since away from the localized regime we have $k_F \ell \gg 1$, the condition becomes approximately $d \lesssim \xi$ (we stress that this condition is for the weak spin-orbit scattering case and thus it is not relevant to the measurements reported in the present work).

For perpendicular field, it was shown in Ref.~\cite{Maki} that for $b\gg \tilde{h}$ we have
\begin{equation}\label{eq:aperp}
    \alpha_\perp = D e H + \frac{\Delta_0}{2b} \tilde{h}^2 = \frac{\Delta_0}{2} \left[ \frac{H}{H_{c2}(0)} + \frac{\tilde{h}^2}{b} \right]
\end{equation}
with $H_{c2}(0)$ of Eq.~(\ref{eq:hc20}).
We see that in both parallel and perpendicular case, the strong spin-orbit scattering adds the term $\Delta_0 \tilde{h}^2/2b$ to the original $\alpha$. There is, however, a qualitative difference in the two cases: while for parallel field the field dependence of $\alpha_\parallel$ is quadratic in the field, for perpendicular field there are both a linear term and a quadratic one. This puts into question the validity of the relationship $H_{c\perp} \propto t$ near $T_c$ as well as that of Eq.~(\ref{WHH}). Clearly, as $b\to \infty$ we can neglect the spin-orbit scattering suppressed Zeeman contribution to $\alpha_\perp$; more precisely, that term can be dropped if $DeH \gg \Delta_0 \tilde{h}^2/2b$. This condition can be written in the form
\begin{equation}\label{eq:perpcond}
  b \gg \left(\frac{\mu_B H_{c2}(0)}{\Delta_0}\right)^2 \frac{H}{H_{c2}(0)}\, .
\end{equation}
This condition should be compared with the assumption $b \gg \tilde{h}= (\mu_B H_{c2}(0)/\Delta_0) H/H_{c2}(0)$ which needs to be satisfied for Eq.~(\ref{eq:aperp}) to be applicable. Now we can distinguish two cases: 1. if $\mu_B H_{c2}(0)/\Delta_0 \lesssim 1$, then when we can use Eq.~(\ref{eq:aperp}), we can always neglect the last term in that equation; then from Eq.~(\ref{eq:atc}) we find $H_{c\perp}(T) \propto t$ near $T_c$ and from Eq.~(\ref{eq:a0ap}) that Eq.~(\ref{WHH}) holds, and moreover $H_{c\perp}(0) = H_{c2}(0)$ of Eq.~(\ref{eq:hc20}). 2. if $\mu_B H_{c2}(0)/\Delta_0 \gg 1$, then one should in general keep the last term in Eq.~(\ref{eq:aperp}), except sufficiently near $T_c$ where $H\to 0$ and hence the condition (\ref{eq:perpcond}) is satisfied; we can then estimate the temperature range over which this happens by using Eq.~(\ref{eq:atc}) to find
\begin{equation}\label{eq:linHval}
  t \ll \frac{\pi^2}{8} e^{-\gamma_E} \left[\frac{\Delta_0}{\mu_B H_{c2}(0)}\right]^2 b\, .
\end{equation}

\subsection{Application to the experimental data}
\label{sec:exp}

The measurements reported in this work were performed near the critical temperature, so for estimating the parameters we will rely on Eq.~(\ref{eq:atc}); the relevant experimental quantities are reported in Table~\ref{tab:data}. From the parallel field data, we can extract the value of $c+1/2b$, see Eq.~(\ref{eq:ap_so}):
\begin{equation}
c+\frac{1}{2b} = - \frac{4}{\pi} \frac{k_B \Delta_0}{\mu_B^2} \left(\frac{dH_{c\parallel}^2}{dT}\right)^{-1}
\end{equation}
We find 0.11 for the 6~nm Re film and 0.033 for the 3~nm one. These estimates bound the values of the spin-orbit scattering parameter, $b_6 \ge 4.5$ and  $b_3 \ge 15$, where we use a subscript to indicate the film's thickness.

\begin{table}
  \centering
 \begin{tabular}{|l|c|c|c|c|}
    \hline
             & $T_c$ & $\Delta_0$ & $-dH_{c2}/dT$ & $-dH^2_{c\parallel}/dT$ \\
             & [$K$] & [meV] & [$T/K$] & [$T^2/K$]\\ \hline
     Re 6 nm & 4.76 & 1.0 & 8.1 & 303 \\
     Re 3 nm & 3.83 & 0.8 & 6.9 & 795 \\
    \hline
  \end{tabular}
  \caption{Summary of the experimentally determined quantities. The gap for the 3~nm film has been estimated from that of the 6~nm one assuming the same gap to critical temperature ratio for the two films.}\label{tab:data}
\end{table}

For the perpendicular field data, let us assume that temperature is sufficiently close to $T_c$ that the condition in Eq.~(\ref{eq:linHval}) is satisfied; we will check later for the consistency of our assumption. Then we can estimate the diffusion constant,
\begin{equation}
  D = -\frac{4}{\pi} \frac{k_B}{e} \left(\frac{dH_{c2}}{dT}\right)^{-1}
\end{equation}
finding $D_6 \simeq 0.14\,$cm$^2$/s and $D_3 \simeq 0.16\,$cm$^2$/s; to these diffusion constants correspond extremely short mean free paths ($v_F = 8\times 10^5\,$m/s) $\ell_6 \simeq 0.05\,$nm and $\ell_3 \simeq 0.06\,$nm.

Putting the diffusion constant together with the value of $\Delta_0$, we arrive at estimates for the coherence length, $\xi_6 \simeq 3.0\,$nm and $\xi_3 \simeq 3.6\,$nm.
Substituting these values of the coherence length in Eq.~(\ref{eq:hc20}) we find $H_{c2}(0) \simeq 36\,$T and 25~T for 6 and 3~nm thickness, respectively. We can now use these estimates together with the bounds on $b$ to check if condition (\ref{eq:linHval}) in fact holds; the lower bounds we find for the right hand side are 0.7 for 6~nm and 3.2 for 3~nm. Looking at Fig.~\ref{OCF}, we see that for both films if we limit the linear fit to fields below 3~T, the temperature range in the experiment is such that condition (\ref{eq:linHval}) is satisfied, meaning we can neglect the quadratic in field Zeeman contribution to $\alpha_\perp$ near $T_c$.

We note, however, that as $T$ decreases towards 0 ($t$ increases to 1), for the 6~nm film the inequality (\ref{eq:linHval}) stops to hold. Nonetheless, let us assume that for both films we can always neglect the quadratic film term in $\alpha_\perp$; then using Eq.~(\ref{eq:a0ap}), or equivalently Eq.~(\ref{WHH}), we obtain $H_{c2}(0) \simeq 27~$T and $18~$T for 6~nm and 3~nm, respectively; the difference between these estimates and the ones above originates from the gap to critical temperature ratio deviating from the BCS value; in Sec.~\ref{sec:results} we reported these more conservative estimates.

So far we have neglected the effect at low temperatures of the quadratic term; it can be included by calculating $H_\perp(0)$ as the solution to the following equation [cf. Eq.~(\ref{eq:aperp})]
\begin{equation}\label{eq:hperp}
  \frac{H}{H_{c2}(0)} + \frac{\tilde{h}^2}{b} = 1
\end{equation}
with $H_{c2}(0)$ obtained as above by neglecting the quadratic term. Here we need the value of parameter $b$; it can be found by calculating $c$ from the definition in Eq.~(\ref{eq:parpar}), using the parameters estimated so far, and then getting $b$ from the known value of $c+1/2b$. For the orbital parameter we obtain $c_6 \simeq 0.037$ and $c_3 \simeq 0.009$, and hence $b_6 \simeq 6.8$ and $b_3 \simeq 20.6$; note that in both cases spin-orbit scattering has a bigger impact on the parallel critical field than the orbital part. The solution to the above equation then gives $H_\perp(0) \simeq 21\,$T for 6~nm and $H_\perp(0) \simeq 17\,$T for 3~nm. As expected, for the thinner film the quadratic Zeeman term does not affect much the estimated $H_\perp(0)$, due to the large spin-orbit scattering $b$ value, while for the thicker film with smaller $b$ the Zeeman effect suppresses $H_\perp(0)$ in comparison with $H_{c2}(0)$.

\subsection{On the validity of Tinkham's formula and the angle-dependent critical temperature}
\label{sec:tink}

As noted already after Eq.~(\ref{eq:aperp}), the pair-breaking strengths $\alpha$ in Eq.~(\ref{eq:aTink}) are modified by the same term $\Delta_0 \tilde{h}^2/2b$ for both parallel and perpendicular direction in the presence of sufficiently strong spin-orbit scattering. This is not surprising: since this term originates from the Zeeman splitting, it does not depend on field direction. Taking this observation into consideration, we deduce that for arbitrary field orientation we have
\begin{equation}\label{eq:atheta}
\alpha(\theta) = D e H \cos\theta + \frac{D}{6\hbar}\left[e d H \sin\theta \right]^2 + \frac{\Delta_0}{2b} \tilde{h}^2
\end{equation}
Since $\tilde{h}^2 = [\tilde{h} \sin \theta]^2 + [\tilde{h}\cos\theta]^2$, it is straightforward to show that when we can neglect the quadratic term in perpendicular field, then Tinkham's formula is still valid. Based on our previous discussion, we therefore expect that formula to hold at temperatures close to $T_c$, while deviations can be present at low temperatures if spin-orbit scattering is not very strong.

\begin{figure}
\includegraphics[width=.44\textwidth]{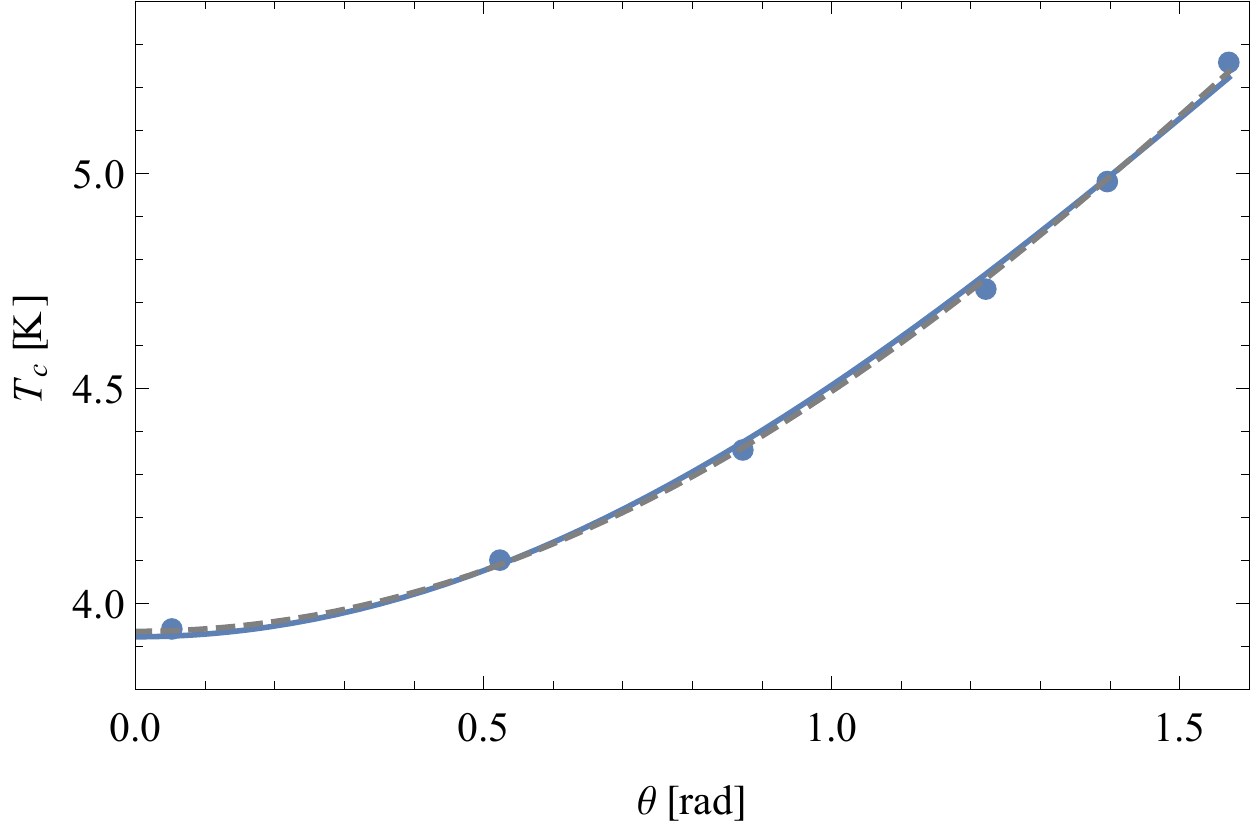}
\caption{Dots: measured critical temperature $T_c$ vs angle $\theta$ from the film normal in a field of magnitude $H=9\,$T. Solid line: best fit line using $\alpha(\theta)$ of Eq.~(\ref{eq:atheta}) for the pair-breaking strength; dashed line: using  $\alpha(\theta)$ of Eq.~(\ref{eq:athetarash}) -- see text.}
\label{fig:tct}
\end{figure}

By substituting Eq.~(\ref{eq:atheta}) into Eq.~(\ref{eq:pbe}), we can interpret the resulting expression as an implicit equation for the critical temperature $T=T_c(\theta)$ as function of angle in a field of fixed magnitude $H$. In Fig.~\ref{fig:tct} we show with dots the result of critical temperature vs angle measurements. The line is a best fit for a 6~nm film where we set $\Delta_0 = 2.4 k_B T_c$ (cf. Sec.~\ref{sec:results}), $b=6.8\times 4.76\,$K$/T_c$ [cf. Eq.~(\ref{eq:parpar})], and treat the
critical temperature $T_c$ and the diffusion constant $D$ as free parameters. We find that both $T_c=5.52\,$K and
$D=0.16\,$cm$^2$/s are higher than the values estimated for the 6~nm film of Sec.~\ref{sec:results}, indicating that the film studied here is
likely less disordered; indeed, the higher $T_c$ is qualitatively consistent with the expectation that increasing disorder suppresses
the critical temperature, as mentioned at the beginning of Sec.~\ref{sec:results}.

\subsection{Rashba effect}

In the discussion of Sec.~\ref{sec:theo}, the enhancement of the parallel critical field above the Pauli limit was ascribed to spin-orbit scattering. Interestingly, a similar enhancement can be caused by a different spin-orbit effect, namely a spin-orbit coupling of the Rashba type. Such couplings can be present in systems lacking inversion symmetry, in  particular due to the presence of interfaces, and break the spin degeneracy. The Rashba coupling is linear in momentum and is characterized by a velocity usually denoted with $\alpha_R/\hbar$ (we use the subscript to avoid confusion with the pair-breaking energy $\alpha$) or, equivalently, by a coupling strength $\Delta_{so} = \alpha_R k_F/\hbar$. Rashba splitting has been observed in ultrathin lead films~\cite{lead}, so it could play a role in our films too.

The effect of the Rahsba coupling on superconductors has been the subject of a number of theoretical works, see~\cite{manuel} and references therein. Here we focus on the regime in which spin-orbit coupling is weak compared to disorder but strong compared to superconductivity,
\begin{equation}\label{eq:Rash_assum}
\frac{\hbar}{\tau} \gg \Delta_{so} \gg \Delta_0 \, ;
\end{equation}
these inequalities imply also that the superconductor is in the disordered -- as opposed to clean -- regime. Under these conditions the pair-breaking energy takes the form~\cite{manuel}
\begin{equation}\label{eq:athetarash}
\begin{split}
\alpha(\theta) = & D e H \cos\theta + \frac{D}{6\hbar}\left[e d H \sin\theta \right]^2  \\ & + \frac{\Delta_0}{2b_R} \tilde{h}^2\left(\sin^2\theta+\frac12 \cos^2\theta\right)
\end{split}
\end{equation}
where
\begin{equation}\label{eq:br}
b_R = \frac{\tau\Delta_{so}^2}{2\hbar\Delta_0}
\end{equation}
Comparing this formula to Eq.~(\ref{eq:atheta}) we see that the difference between Rahsba spin-orbit coupling and spin-orbit scattering manifests itself via a different angular dependence of the last term.

The considerations in Secs.~\ref{sec:theo} and \ref{sec:exp} can be repeated with the simple replacements $b\to b_R$ when examining parallel field and $b\to 2b_R$ in the perpendicular configuration. Since the values for parameter $b$ were estimated from data near $T_c$ where the contributions of spin-orbit effects to the critical field can be neglected, we have $b_{R,6}= b_6$ and $b_{R,3}=b_3$.
The only modifications in our estimates take place in the calculations of the spin-orbit suppression of the zero-temperature perpendicular critical field $H_\perp(0)$ presented after Eq.~(\ref{eq:hperp}), which now give $23\,$T and $17.6\,$T for the 6~nm and 3~nm film, respectively; note that the additional factor of 2 in the replacement for the perpendicular field case reduces the impact of Rashba spin-orbit coupling on the perpendicular critical field as compared to spin-orbit scattering. Moreover, using Eq.~(\ref{eq:br}) we can estimate $\Delta_{so,6}\simeq 0.38\,$eV and $\Delta_{so,3} \simeq 0.54\,$eV, values that satisfy the conditions in Eq.~(\ref{eq:Rash_assum}). Correspondingly, we find $\alpha_{R,6} \simeq 0.26\,$eV\r{A} and $\alpha_{R,3} \simeq0.37\,$eV\r{A}. The finding $\alpha_{R,6} < \alpha_{R,3}$ is qualitatively in agreement with the expectation of stronger Rashba effect in thinner films. The magnitude of the Rashba parameters seems reasonable: while they are larger than what was found in lead films~\cite{lead} and some semiconductors, they are smaller than those found in other semiconductors, metallic surfaces, and topological insulators~\cite{nmat}.

Finally, using Eq.~(\ref{eq:athetarash}) we can repeat the fitting of the critical temperature vs angle data in Fig.~\ref{fig:tct}, obtaining the dashed line with fit parameters $T_c = 5.54\,$K and $D=0.17\,$cm$^2$/s. The fit is marginally better (smaller sum of squared residuals) than that obtained using Eq.~(\ref{eq:hperp}). The analysis in this section shows that the Rashba effect could be a viable explanation of the spin-orbit enhancement of the parallel critical field in our films; however, we cannot exclude spin-orbit scattering based on our measurements.

\section{Critical field transitions}
\label{app:cft}

\begin{figure}[!tb]
\begin{flushleft}
\includegraphics[width=.44\textwidth]{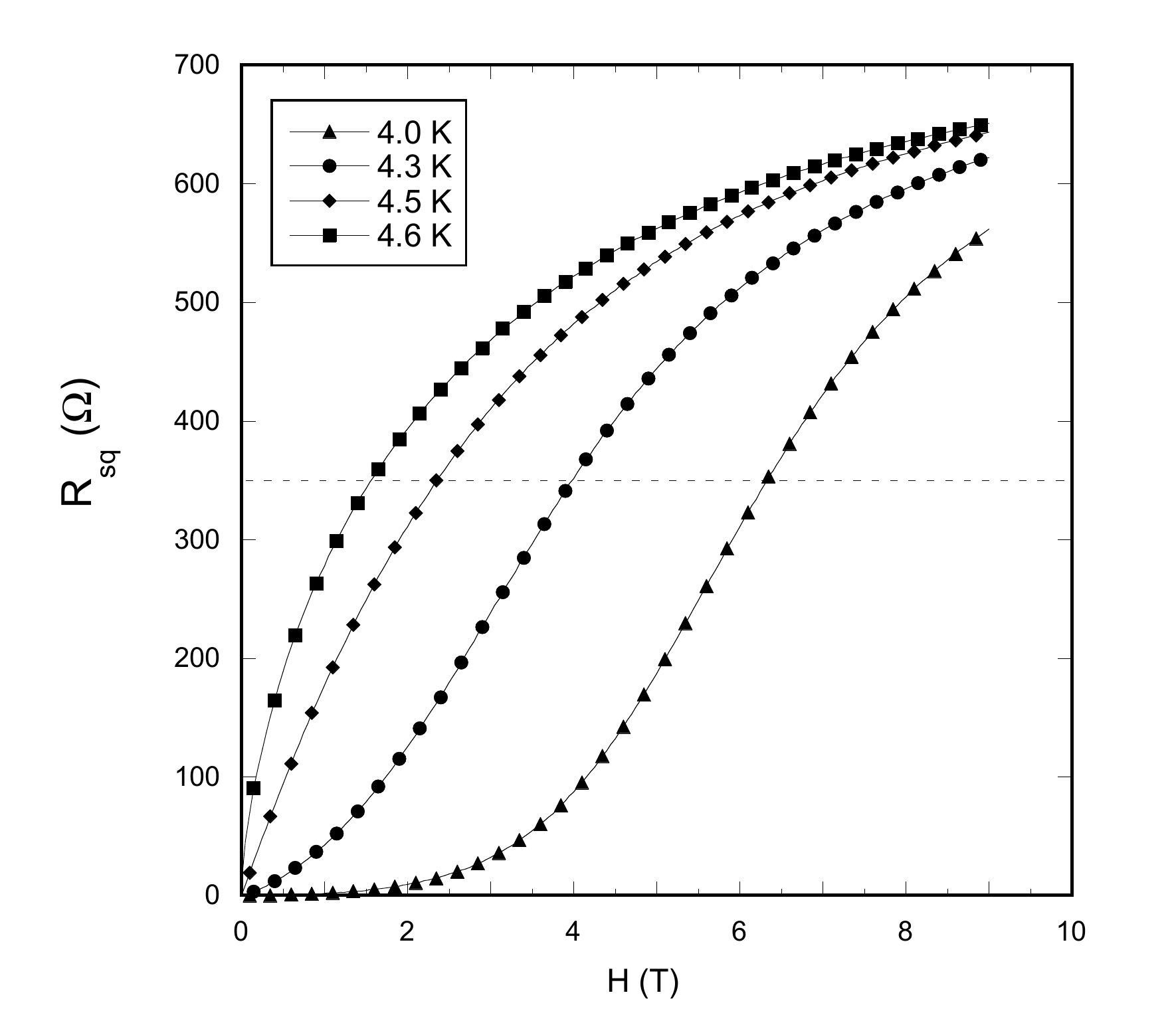}\end{flushleft}
\caption{Sheet resistance as a function of perpendicular magnetic field at several temperatures near $T_c$ for a 6 nm Re film.   The horizontal dashed line represents $R_n/2$ which defines the critical field.}
\label{R-H-T-60}
\end{figure}

\begin{figure}[!bt]
\begin{flushleft}
\includegraphics[width=.44\textwidth]{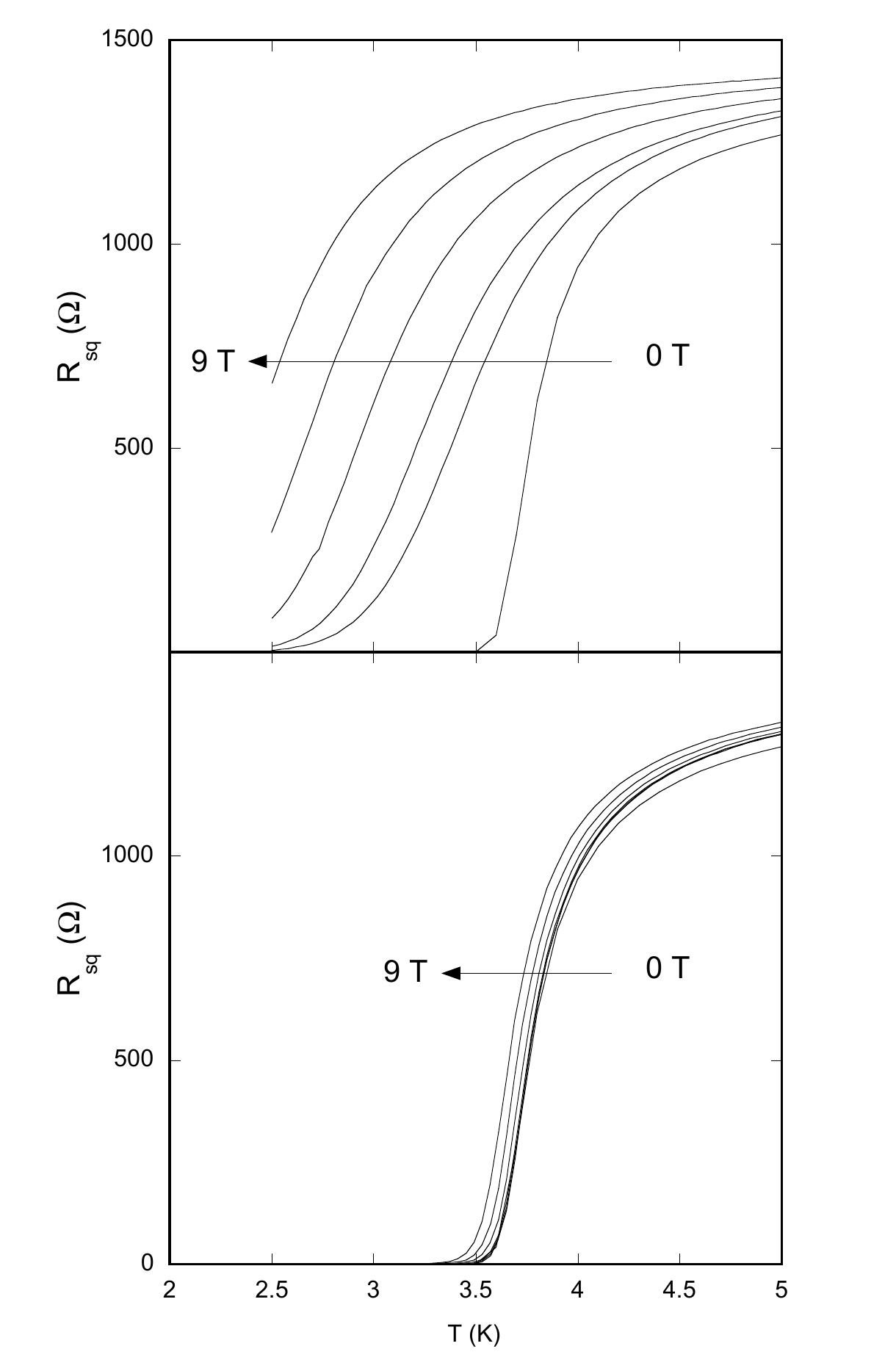}\end{flushleft}
\caption{Superconducting transitions of a 3 nm Re film in the presence of a range of applied fields.  Upper panel: sheet resistance as a function of temperature with applied perpendicular fields of 0, 2, 3, 5, 7, and 9 T. Lower panel: superconducting transitions in applied parallel fields of 0, 1, 2, 3, 5, 7, and 9 T.}
\label{R-T-H-30}
\end{figure}

In order to estimate the $T=0$ critical fields, we have employed Eqs.~(\ref{WHH}) and (\ref{Hc||}).   The equations require that we measure temperature dependence of the critical field near $T_c$.  This can be done by directly scanning the field at temperatures near $T_c$ or, alternatively, by measuring the transition temperature in the presence of a static applied field.  In either case the transition was define by the temperature/field at which the resistance reached $1/2$ of its normal state value $R_n$. These two methods gave equivalent results but sweeping temperature in a constant magnetic field proved to be more expedient.  The first method is illustrated in Fig.~\ref{R-H-T-60} were several isothermal critical field transitions are measured at temperatures near $T_c$ in a 6 nm Re film.  The second method is represented by the data in Fig.~\ref{R-T-H-30} where temperature is swept through the transition in constant perpendicular and parallel magnetic fields.


\begin{thebibliography}{30}
\bibitem{UTe2-b}  The extraordinarily high critical field in UTe$_2$ is extremely sensitive to field orientation.
\end{thebibliography}
\end{document}